\documentclass[aip,reprint,graphicx,floatfix]{revtex4-1}

\usepackage{graphicx}
\usepackage{amsmath}

\begin{document}


\title{Critical behavior in a chiral molecular model}
\author{Pablo M. Piaggi}
\affiliation{Department of Chemistry, Princeton University, Princeton, NJ 08544, USA}
\author{Roberto Car}
\affiliation{Department of Chemistry, Princeton University, Princeton, NJ 08544, USA}
\affiliation{Department of Physics, Princeton University, Princeton, NJ 08544, USA}
\author{Frank H. Stillinger}
\affiliation{Department of Chemistry, Princeton University, Princeton, NJ 08544, USA}
\author{Pablo G. Debenedetti}
 \affiliation{Department of Chemical and Biological Engineering, Princeton University, Princeton, NJ 08544, USA}
 \email{pdebene@princeton.edu}

\date{\today}

\begin{abstract}
Understanding the condensed-phase behavior of chiral molecules is important in biology, as well as in a range of technological applications, such as the manufacture of pharmaceuticals.
Here, we use molecular dynamics simulations to study a chiral four-site molecular model that exhibits a second-order symmetry-breaking phase transition from a supercritical racemic liquid, into subcritical D-rich and L-rich liquids.
We determine the infinite-size critical temperature using the fourth-order Binder cumulant, and we show that the finite-size scaling behavior of the order parameter is compatible with the 3D Ising universality class.
We also study the spontaneous D-rich to L-rich transition at a slightly subcritical temperature $T\approx0.985 T_c$ and our findings indicate that the free energy barrier for this transformation increases with system size as $N^{2/3}$ where $N$ is the number of molecules, consistent with a surface-dominated phenomenon.
The critical behavior observed herein suggests a mechanism for chirality selection in which a liquid of chiral molecules spontaneously forms a phase enriched in one of the two enantiomers as the temperature is lowered below the critical point.
Furthermore, the increasing free energy barrier with system size indicates that fluctuations between the L-rich and D-rich phases are suppressed as the size of the system increases, trapping it in one of the two enantiomerically-enriched phases.
Such a process could provide the basis for an alternative explanation for the origin of biological homochirality.
We also conjecture the possibility of observing nucleation at subcritical temperatures under the action of a suitable chiral external field.
\end{abstract}

\maketitle

\raggedbottom

\section{Introduction}

Fundamental building blocks involved in the complex machinery of biological cells exist as a single enantiomer, one of the two possible isomers of a chiral molecule.
For instance, naturally-occurring amino acids are left-handed while sugars are right-handed\cite{blackmond2010origin,ball2007giving}.
The emergence of this phenomenon, known as biological homochirality, remains incompletely understood\cite{blackmond2010origin}.
Several theories have attempted to shed light on the origin of biological homochirality, including the explanation that an enantiomer may act as catalyst for its own formation\cite{frank1953spontaneous,blackmond2004asymmetric}.
Another theory proposes that chiral symmetry breaking can emerge from the equilibrium solid-liquid phase behavior of amino acids in solution\cite{klussmann2006thermodynamic,klussmann2006rationalization}.
It has also been suggested that parity violation could account for very small energy differences between enantiomers and thus give rise to a preferred chirality\cite{quack2002important}.
Biological homochirality is also of importance in the food, cosmetic, and pharmaceutical industries, as enantiomers can interact in dramatically different ways with receptors in the human body\cite{sanchez2004escitalopram}.

Simulations can provide atomistic-level insight into the behavior of molecular systems, including chiral molecules in particular.
Latinwo et al.~\cite{latinwo2016molecular} introduced a molecular model for a chiral tetramer suitable for molecular dynamics (MD) simulations.
The model is able to switch between two chiral conformers and describe intermolecular interactions in the condensed phase.
A key feature of this model is a chiral renormalization factor, which can be tailored to enhance homochiral or heterochiral short-range interactions.
The thermodynamics of this model were recently investigated by Wang et al.~\cite{wang2022fluid}, who showed that this model exhibits spontaneous symmetry breaking below a critical temperature when homochiral interactions are favored.
At these conditions, two symmetry-equivalent phases can form: a D-rich liquid and an L-rich liquid.
Spontaneous symmetry breaking in a liquid of chiral molecules has also been observed experimentally by Dressel et al.\cite{dressel2014chiral}

Here, we revisit the chiral tetramer model, and investigate its critical behavior and liquid-liquid transition.
Our findings confirm the existence of a critical point in this model, and provide insight into the universality class it belongs to.
We also study the kinetics of the liquid-liquid transition and suggest a scaling law for the free energy barrier with system size.
Finally, we analyze the cluster size distribution during the liquid-liquid transformation in order to better understand the mechanism of this transition.

\begin{figure*}
\includegraphics[width=0.95\textwidth]{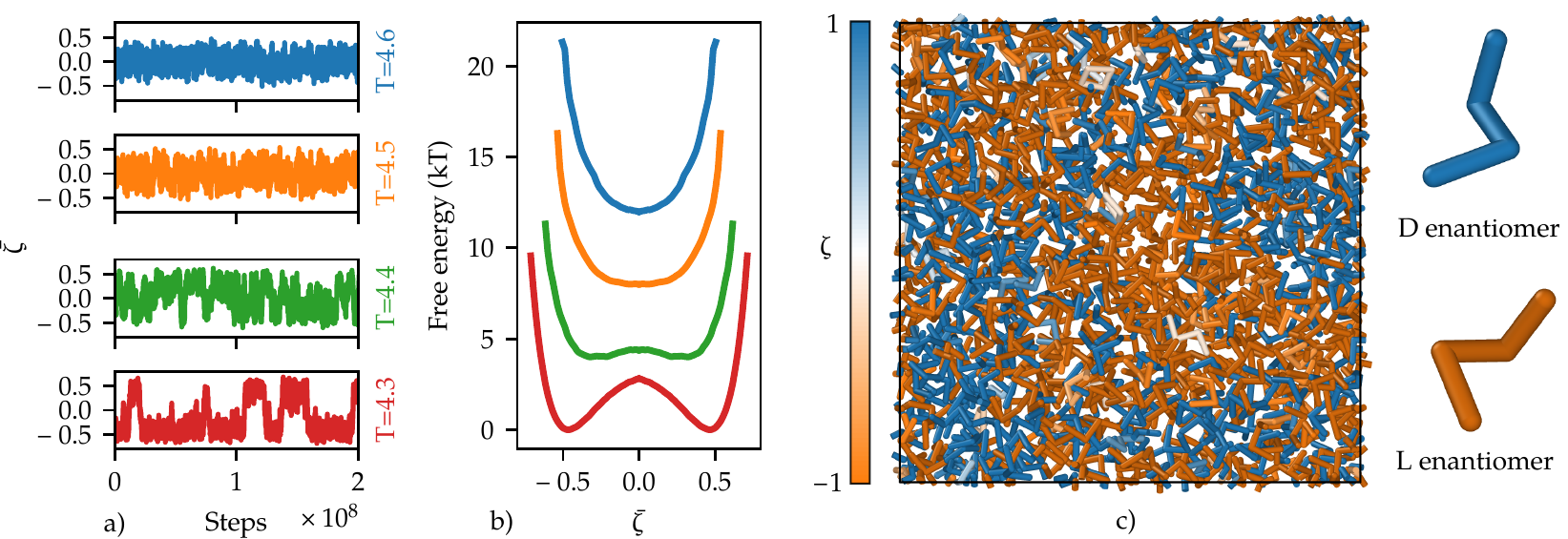}
\caption{\label{fig:Fig1} Liquid-liquid interconversion in the chiral tetramer model with system size $N=1000$. a) Mean chirality $\bar\zeta$ vs.~simulation steps for temperatures $T=4.3$, 4.4, 4.5, and 4.6. b) Free energy as a function of the order parameter $\bar\zeta$. Curves are color-coded to match temperatures in panel a). We have imposed even parity of the free energy curves due to symmetry considerations. c) Snapshot of a configuration at $T=4.6$. Molecules are colored according to their chirality $\zeta$. Enantiomers D and L are also shown.}
\end{figure*}

\section{Methods}

\subsection{Molecular Chiral Model}

The model introduced by Latinwo et al.~\cite{latinwo2016molecular} considers four-bead tetramers able to switch between two chiral conformations (see FIG.~\ref{fig:Fig1}c).
The potential energy within this model is:
\begin{equation}
U(\mathbf{R}) = U_{\mathrm{intra}}(\mathbf{R}) + U_{\mathrm{inter}}(\mathbf{R}),
\label{eq:pe}
\end{equation}
where $U_{\mathrm{intra}}$ and $U_{\mathrm{inter}}$ are intra- and inter-molecular interactions, respectively, and $\mathbf{R}$ are the atomic coordinates.
The intramolecular interactions in a system of $N$ molecules are given by:
\begin{align}
    U_{\mathrm{intra}}(\mathbf{R}) = \sum_{i=1}^N \left ( \sum_{j=1}^3 \frac{k_b}{2}(|\mathbf{r}^i_j-\mathbf{r}^i_{j+1}|-b)^2 + \nonumber \right. \\
    \left. \sum_{k=1}^2 \frac{k_a}{2}\left( \theta^i_k - \frac{\pi}{2} \right) + k_d \cos^2\phi^i \right )
    \label{eq:pe_intra}
\end{align}
where $\mathbf{r}^i_j$ is the position of bead $j$ ($j=1,2,3,4$) of the $i$-th molecule, $\theta^i_k$ with $k=1,2$ are the two bond angles in the $i$-th molecule, and $\phi^i$ is the dihedral angle of the $i$-th molecule.
The strength of the intramolecular interactions is controlled by the force constants $k_b$, $k_a$, and $k_d$ for the bond length, angle, and dihedral.
The equilibrium molecular geometry is given by the bond length $b$, the angles $\pi/2$, and the dihedral $\pm\pi/2$.
The two chiral molecular conformers are characterized by dihedral angles $+\pi/2$ and $-\pi/2$.

The intermolecular interactions in this model are given by:
\begin{equation}
    U_{\mathrm{inter}}(\mathbf{R}) = \sum_{i=1}^N \sum_{j=i+1}^{N} \sum_{l=1}^4 \sum_{m=1}^4 \epsilon(\mathbf{R}^i,\mathbf{R}^j) \: v_{LJ}\left ( \frac{|\mathbf{r}^i_l-\mathbf{r}^j_m|}{\sigma} \right )  
\end{equation}
where indices $i$ and $j$ refer to molecules, and $l$ and $m$ to the four beads in each molecule.
$v_{LJ}$ is the smooth force variant\cite{toxvaerd2011communication} of the Lennard-Jones potential, i.e.:
\begin{equation}
  v_{LJ}(r) = 
  \begin{cases}
  \phi(r)-\phi(r_c)-(r-r_c)\phi'(r_c) \quad \mathrm{if} \: r \leq r_c \\
  0  \quad \mathrm{if} \: r>r_c
  \end{cases}
\end{equation}
where $\phi(r)=4(r^{-12}-r^{-6})$ is the standard Lennard-Jones potential, $r_c$ is the cutoff for the $v_{LJ}(r)$ potential, and $\phi'(r)$ is the first derivative of $\phi(r)$ with respect to $r$.
A key feature in this model is the chiral renormalization factor $\epsilon(\mathbf{R}^i,\mathbf{R}^j)$, which controls the strength of the intermolecular interactions, and depends on the chiral state of molecules $i$ and $j$ defined through vectors $\mathbf{R}^i=\{\mathbf{r}^i_1,\mathbf{r}^i_2,\mathbf{r}^i_3,\mathbf{r}^i_4\}$ and $\mathbf{R}^j=\{\mathbf{r}^j_1,\mathbf{r}^j_2,\mathbf{r}^j_3,\mathbf{r}^j_4\}$ that contain the coordinates of the four sites of the $i$-th and $j$-th molecules, respectively.
The functional form of the renormalization factor is:
\begin{equation}
\epsilon(\mathbf{R}^i,\mathbf{R}^j) = \epsilon_0 [1+\lambda \zeta(\mathbf{R}^i) \zeta(\mathbf{R}^j)]
\end{equation}
where $-1<\zeta<1$ represents the chirality of each molecule.
The behavior of the model can be tuned via the parameter $\lambda$ which introduces a preference for homochiral intermolecular interactions if it is positive, and heterochiral interactions if it is negative.
For the four-site model considered here, the chirality of each molecule can be calculated as:
\begin{equation}
    \zeta(\mathbf{R}^i)=-\frac{\mathbf{r}^i_{12} \cdot (\mathbf{r}^i_{23} \times \mathbf{r}^i_{34}) }{|\mathbf{r}^i_{12}| \: |\mathbf{r}^i_{23}| \: |\mathbf{r}^i_{34}|}
    \label{eq:chirality_zeta}
\end{equation}
where $\mathbf{r}^i_{jk}=\mathbf{r}^i_{j}-\mathbf{r}^i_{k}$.
With the definition in Eq.~\eqref{eq:chirality_zeta}, $\zeta=1$ for D enantiomers and $\zeta=-1$ for L enantiomers.
We note that the mathematical definition of $\zeta(\mathbf{R}^i)$ in Eq.~\eqref{eq:chirality_zeta} captures the essence of chirality, i.e., $\zeta(\mathbf{R}^i)$ is by construction a pseudo-scalar and thus changes sign under a parity transformation (reflection with respect to a plane).

\subsection{Molecular Dynamics Simulations}

We performed all simulations using the MD engine LAMMPS\cite{plimpton1995fast} (version February 1, 2014) and an implementation of the potential energy of the chiral model based on ref.~\citenum{petsev2021chiral-dataspace}.
The intermolecular term of the potential energy, $U_{\mathrm{inter}}(\mathbf{R})$, gives rise to an eight-body force involving all four sites of pairs of tetramer molecules.
This force was calculated as described by Petsev et al.\cite{petsev2021effect}

We carried out the simulations in the canonical ensemble with density 0.11 molecules$/\sigma^3$.
We controlled the temperature constant using a Nosé-Hoover thermostat\cite{nose1984unified,hoover1985canonical} and we employed a time step of $10^{-3}$ (reduced units are used unless otherwise specified).
The parameters of the chiral molecular model were $k_b=8003$, $k_a=643.7$, $k_d=17.86$, $b=1.0583$, $\sigma=1$, $\epsilon_0=1$, $r_c=4.0$, and the mass of the tetramer beads was set to $m=1$. 
Following ref.~\citenum{wang2022fluid} we chose $\lambda=0.5$ in order to favor homochiral interactions.
We rebuilt neighbor lists\cite{FrenkelBook} every 10 steps with cutoff $4\sigma$ and we used a neighbor-list skin of $0.3\sigma$.
For this thin neighbor-list skin, we observed a small drift of the total energy in NVE simulations over long simulation times.
To avoid artifacts, we used a neighbor-list skin of $1.5\sigma$ for simulations longer than $5 \times 10^8$ steps.
We recommend a neighbor-list skin of $1.5\sigma$ for future simulations.

\subsection{Clustering}

\label{sec:clustering}

We performed clustering using a distance-based criterion wherein two beads are considered to be adjacent if 1) their distance is below $1.5\sigma$ and 2) they have the same chirality (both molecules satisfy $\zeta>0$ or $\zeta<0$).
Based on this adjacency criterion we subsequently find the connected components (clusters) using the analysis software Freud 2.8.0.\cite{Ramasubramani20}
Since the equilibrium distance between beads within a molecule is around $\sigma$, we can calculate the number of connected molecules $n_c$ within a cluster as the total number of connected beads divided by the number of beads in a molecule, i.e., four.
This algorithm could fail to recognize that two beads in a molecule are connected if their distance were greater than 1.5$\sigma$.
However, this event has negligible probability and was not observed in our simulations.
An alternative clustering strategy could have employed the centers of mass of the tetramers.

\section{Results}

\subsection{Critical behavior}

We first investigate the behavior of the model at constant density and varying temperature, for a system of 1000 tetramer molecules.
In FIG.~\ref{fig:Fig1}a we show the mean chirality, defined as $\bar\zeta = (1/N) \sum_{i=1}^N \zeta(\mathbf{R}^i)$, as a function of the simulation steps for temperatures $T$ equal to 4.3, 4.4, 4.5, and 4.6.
The mean chirality $\bar\zeta$ is a natural order parameter for this model.
At $T=4.5$ and $T=4.6$, $\bar\zeta$ fluctuates around 0, indicating the presence of a racemic mixture at these conditions.
If the temperature is lowered to $T=4.4$, fluctuations in $\bar\zeta$ increase significantly and span a relatively large range $\bar\zeta \in (-0.5,0.5)$.
At an even lower temperature $T=4.3$, $\bar\zeta$ shows clear bistable behavior and alternates between a D-rich phase and a L-rich phase.

In order to provide further insight into this behavior, we calculate the free energy as a function of $\bar\zeta$ which we define as $F(\bar\zeta)=- k_B T \ln P(\bar\zeta)$ where $P(\bar\zeta)$ is the probability of observing a given value of $\bar\zeta$ and $k_B$ is the Boltzmann constant.
$F(\bar\zeta)$ is shown in FIG.~\ref{fig:Fig1}b for the four temperatures described above.
As expected, for $T\gtrsim 4.5$ the free energy profiles show a single minimum centered at $\bar\zeta=0$.
At $T=4.4$ the free energy profile is relatively flat and starts to develop two minima.
Finally, at $T=4.3$ the symmetry breaking becomes evident and $F(\bar\zeta)$ shows two symmetric, well-defined minima.
Overall, the results are characteristic of a second-order, symmetry-breaking phase transition with a critical point slightly above $T=4.4$.
As a result of the symmetry breaking, below the critical point enantiomers can be classified as minority and majority enantiomers, depending on which of the two enantiomers is predominant.
In FIG.~\ref{fig:Fig1}c, we show a snapshot of a configuration at supercritical temperature $T=4.6$ that corresponds to a racemic mixture.

\subsection{Infinite-size critical temperature and finite-size scaling}

\begin{figure}
\includegraphics[width=\columnwidth]{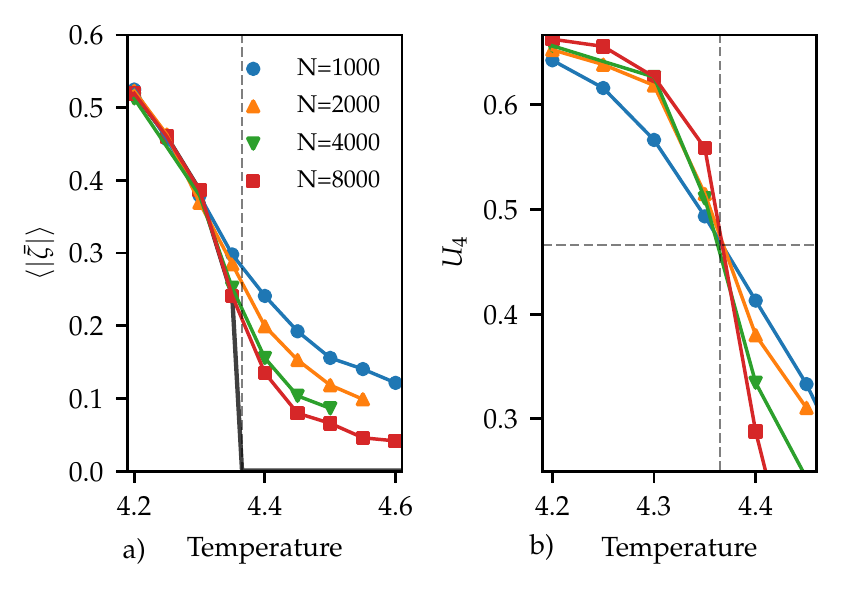}
\caption{\label{fig:Fig2} Finite-size effects in critical behavior of the chiral tetramer. a) Ensemble average $\langle |\bar\zeta| \rangle $ of the mean chirality $\bar\zeta$ vs. temperature for systems sizes $N=1000$, 2000, 4000, and 8000. The solid black line is a linear interpolation to the expected behavior of $\langle |\bar\zeta| \rangle $ for $N \to\infty$, namely, $\langle |\bar\zeta| \rangle =0$ for $T \geq T_c$, with $T_c$ the infinite-size critical temperature. b) Fourth-order Binder cumulant ($U_4$) vs.~temperature. The fixed point $U_4^*$ is marked with a horizontal dashed line. The infinite-system-size critical temperature is shown in a) and b) with a vertical dashed line.}
\end{figure}

The results described above correspond to a system of 1000 molecules.
Thus, they can be affected by finite-size effects.
For this reason, we also studied systems of 2000, 4000, and 8000 molecules.
In FIG.~\ref{fig:Fig2}a, we show the ensemble average of $|\bar\zeta|$ (denoted by $\langle |\bar\zeta| \rangle $) as a function of temperature, for different system sizes.
$\langle |\bar\zeta| \rangle $ vanishes at high temperatures and takes on a finite value at low temperatures.
As expected for a second order phase transition, the crossover from the high-temperature to the low-temperature phase is smoother for smaller system sizes.
The critical temperature in the thermodynamic limit can be estimated from finite system simulations using the fourth-order Binder cumulant \cite{binder1981finite}:
\begin{equation}
    U_4=1-\frac{\langle \bar\zeta^4 \rangle }{3\langle \bar\zeta^2 \rangle^2}
\end{equation}
which has a fixed point $U_4^*$ at the infinite-size critical temperature for any system size.
In FIG.~\ref{fig:Fig2}b we show $U_4$ vs temperature for different system sizes.
The critical temperature can be determined by the fixed point at which these curves cross each other.
From this analysis we determine that $T_c \approx 4.365$ in the thermodynamic limit.
Furthermore, $U_4$ at the fixed point ($U_4^*$) depends only on the universality class.
Here, we found $U_4^*\approx0.46$ which is in good agreement with the value for the 3D Ising universality class $U_4^*=0.466(1)$\cite{ferrenberg2018pushing}.


\begin{figure}
\includegraphics[width=\columnwidth]{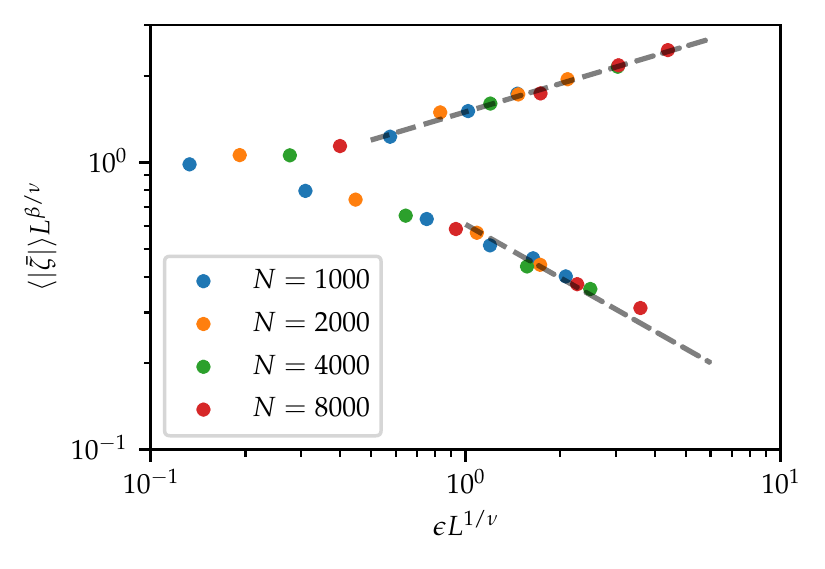}
\caption{\label{fig:fss} Finite-size scaling of the order parameter $\langle |\bar \zeta| \rangle$. $L$ is a distance such that $L \propto N^{1/3}$, $\beta$ and $\nu$ are critical exponents, and $\epsilon=(T-T_c)/T_c$. The dashed lines illustrate the power law behavior of $\langle |\bar \zeta| \rangle \: L^{\beta/\nu}$ at temperatures well above and well below the critical temperature. The slope of the upper branch of the curve is $\beta$ and the slope of the lower branch is $\beta-3 \nu/2$.}
\end{figure}

We also study the finite-size scaling of the order parameter $\bar\zeta$ in order to provide further evidence of this system's universality class.
The order parameter in a second order phase transition follows the following finite-size scaling\cite{landau2021guide}:
\begin{equation}
M \: L^{\beta/\nu} = \mathcal{M}^0 (\epsilon L^{1/\nu})
\label{eq:fss}
\end{equation}
where $M$ is the order parameter, $L$ is the simulation box side, $\beta$ and $\nu$ are critical exponents, $\mathcal{M}_0$ is a universal function, and $\epsilon=|T-T_c|/T_c$.
We thus plot $\langle |\bar \zeta| \rangle \: L^{\beta/\nu}$ vs.~$\epsilon L^{1/\nu}$ using our simulation data at different temperatures and system sizes.
We choose the exponents $\beta=0.326419(3)$ and $\nu=0.629971(4)$ of the 3D Ising universality class\cite{el2014solving}.
Our results, depicted in FIG.~\ref{fig:fss}, show that data for all system sizes follow the universal behavior predicted by the finite-size scaling hypothesis in Eq.~\eqref{eq:fss}.
These results lend strong support to the hypothesis that the chiral tetramer system belongs to the 3D Ising universality class, since other choices of exponents $\beta$ and $\nu$ do not lead to correct finite-size scaling.

The dashed lines in FIG.~\ref{fig:fss} illustrate the power law behavior of $\langle |\bar \zeta| \rangle \: L^{\beta/\nu}$ at temperatures well above and well below the critical temperature.
In the low-temperature regime, $\langle |\bar \zeta| \rangle$ is constant with system size and $\langle |\bar \zeta| \rangle L^{\beta/\nu} \propto L^{\beta/\nu} = x^\beta$ with $x=L^{1/\nu}$.
Thus, the slope of the upper branch of the curve in FIG.~\ref{fig:fss} is $\beta$.
On the other hand, in the high-temperature regime $\langle |\bar \zeta| \rangle \propto N^{-1/2}$ and $\langle |\bar \zeta| \rangle L^{\beta/\nu} \propto L^{\beta/\nu-3/2} = x^{\beta-3 \nu/2}$. Therefore, $\beta-3 \nu/2$ is the slope of the lower branch of the curve in FIG.~\ref{fig:fss}.
The data in FIG.~\ref{fig:fss} indeed supports the slopes derived above.
We note that the slopes for the power law behavior were obtained for $\langle |\bar \zeta| \rangle L^{\beta/\nu}$ as a function of $L^{1/\nu}$, while in FIG.~\ref{fig:fss} we plot $\langle |\bar \zeta| \rangle L^{\beta/\nu}$ vs. $\epsilon L^{1/\nu}$ (instead of $L^{1/\nu}$).
This approximation is valid if $\log(L^{1/\nu})/\log(\epsilon)>>1$ which is satisfied by our data for a sufficiently large $\epsilon L^{1/\nu}$.


\begin{figure*}
\includegraphics[width=\textwidth]{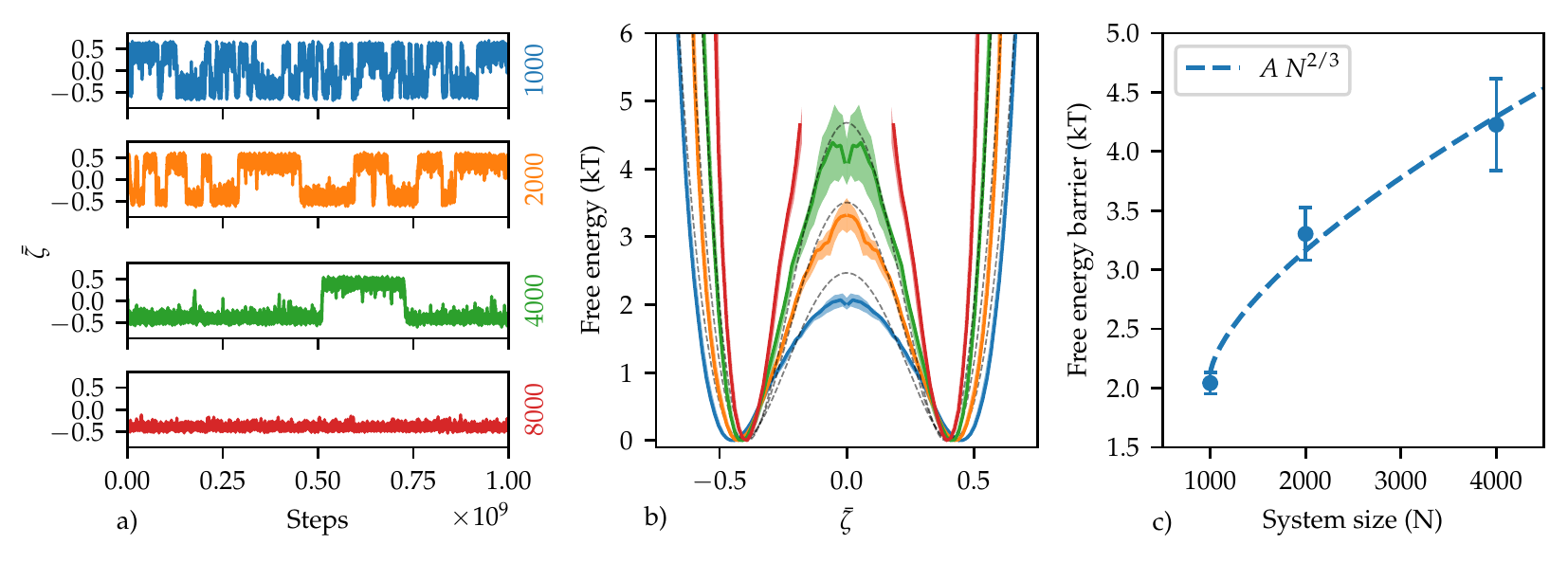}
\caption{\label{fig:kinetics} Liquid-liquid interconversion at subcritical temperature $T=4.3$. a) Mean chirality $\bar\zeta$ vs.~simulation steps for system sizes $N=1000$, 2000, 4000, and 8000. b) Free energy as a function of $\bar\zeta$. We have imposed even parity of the free energy curves due to symmetry considerations. The shaded area is the one standard deviation error calculated with four-fold block averaging. The dashed lines are fits to the Landau free energy model. c) Free energy barriers $\Delta F^\ddag$ with error bars representing the standard deviation of the mean calculated using four-fold block averages. A fit to the expression $\Delta F^\ddag \propto N^{2/3}$, which corresponds to a surface-dominated barrier, is shown with a dashed line.}
\end{figure*}

\subsection{Kinetics of the liquid-liquid interconversion}

We now analyze the kinetics of the transition between the $D$-rich and the $L$-rich phase at subcritical temperature $T=4.3$.
In FIG.~\ref{fig:kinetics}a we show $\bar\zeta$ as a function of the simulation steps for different system sizes.
The results clearly indicate that the interconversion between these phases becomes more infrequent for larger systems.
We also calculated the free energies as a function of the order parameter $\bar\zeta$ and we show them in FIG.~\ref{fig:kinetics}b.
The free energy near $\bar\zeta \approx 0$ is not well resolved for the largest system ($N=8000$) due to poor sampling.
The free energies in FIG.~\ref{fig:kinetics}b can be fit to a Landau model\cite{LandauBookStatisticalPhysics}:
\begin{equation}
    F(\bar\zeta)= A (T-T_c) \bar\zeta^2 + B \bar\zeta^4
    \label{eq:landau}
\end{equation}
where $A>0$ and $B>0$ have to be determined.
The Landau model fits the simulation results well (see FIG.~\ref{fig:kinetics}b), and the parameters $A$, $B$, and $T_c$ determined for each system size are shown in Table 1.
\begin{table}
\caption{\label{tab:table1} Parameters in the Landau model (Eq.~\eqref{eq:landau}) $A$, $B$, and $T_c$ for different system sizes $N$ at temperature $T=4.3$. Errors were calculated as the standard deviation of the mean using four-fold block averages.}
\begin{ruledtabular}
\begin{tabular}{ccccc}
         $N$ & $A$ & $B$ & $T_c$ \\ 
         \hline
1000 & 467(14) & 79(1) &  4.36(1) \\
2000 & 704(26) & 135(3)  &  4.36(1)  \\
4000 & 1010(10) & 209(9)  &  4.36(1)  \\
\end{tabular}
\end{ruledtabular}
\end{table}

The results in FIG.~\ref{fig:kinetics}b clearly show that the free energy barrier for the transition becomes progressively larger as a function of system size.
The free energy barriers $\Delta F^\ddag$ are plotted in FIG.~\ref{fig:kinetics}c, and appear to scale with system size as $\Delta F^\ddag \propto N^{2/3}$, compatible with a surface-dominated phenomenon.
The same scaling ($\Delta F^\ddag \propto N^{2/3}$) is well-known for the Ising model\cite{chandler1987introduction}.
Note also that the barrier $\Delta F^\ddag$ diverges in the thermodynamic limit, signaling ergodicity breaking, and thus liquid-liquid interconversion is a finite-size effect.
The interconversion time increasing with system size can be understood by noting that when transitioning from an L-rich to a D-rich phase (or viceversa), the system must necessarily traverse the condition $\bar\zeta = 0$.
This corresponds to the maximum possible number of energetically unfavorable heterochiral contacts ($\lambda > 0$), and one expects this effective barrier to scale as $N^{2/3}$.
This is consistent with the behavior shown in FIG.~\ref{fig:kinetics}c.

\subsection{Nucleation}

The dependence of the interconversion rate with system size shown in FIG.~\ref{fig:kinetics}a is, of course, at odds with the predictions of nucleation theory.
The nucleation rate is defined as $J=1/(\tau V)$ where $\tau$ is the average time to form a critical cluster in a volume $V$.
Within classical nucleation theory, the rate is:
\begin{equation}
    J \propto \exp \left ( \frac{-\Delta F}{k_B T} \right )
    \label{eq:nuc_rate}
\end{equation}
where $\Delta F$ is a microscopic barrier, independent of system size.
Thus, in nucleation phenomena one expects to observe a faster transition (smaller $\tau$) as the size of the system increases, while in FIG.~\ref{fig:kinetics}a we observe the opposite trend.

The absence of nucleation in the D-rich to L-rich phase transformation is not surprising.
The driving force for nucleation is the difference in chemical potential between the mother phase and the new phase.
In this case, the D-rich phase and the L-rich phase have the same chemical potential $\mu_L=\mu_D$ and thus there is no driving force for nucleation.
The equality of the chemical potentials follows from the fact that the energy in this model is invariant under a parity transformation.
Thus, it is impossible to trigger a phase transformation between the D-rich and the L-rich liquids through thermodynamic means, i.e., via changes in temperature and pressure.
One may speculate that in a realistic chiral molecule, parity violation could lead to  $\mu_L \neq \mu_D$.
However, the difference in chemical potential would likely be too small to promote nucleation.

The behavior described above for the chiral molecular model is similar to that of the Ising model in the absence of an external field\cite{landau2021guide}.
In analogy to the Ising model, we conjecture that one may introduce an external field $h$ that couples to the order parameter $\bar\zeta$.
The potential energy in the presence of an external field can be written as:
\begin{equation}
U'(\mathbf{R}) = U(\mathbf{R}) - N \: \bar\zeta \: h,
\label{eq:pe_field}
\end{equation}
where $U(\mathbf{R})$ is defined in Eq.~\eqref{eq:pe}, and $h$ is a scalar field.
The added term $N \bar\zeta \: h$ is chiral (a pseudoscalar) and thus creates a chemical potential imbalance $\mu_L \neq \mu_D$ for $h \neq 0$.
If a suitable chiral external field can be devised, it should be possible to trigger the spontaneous transformation between the D-rich and the L-rich liquids.
Furthermore, we expect the transformation to proceed through nucleation and growth.

\subsection{Phase diagram at constant volume and number of molecules}

\begin{figure}
\includegraphics[width=\columnwidth]{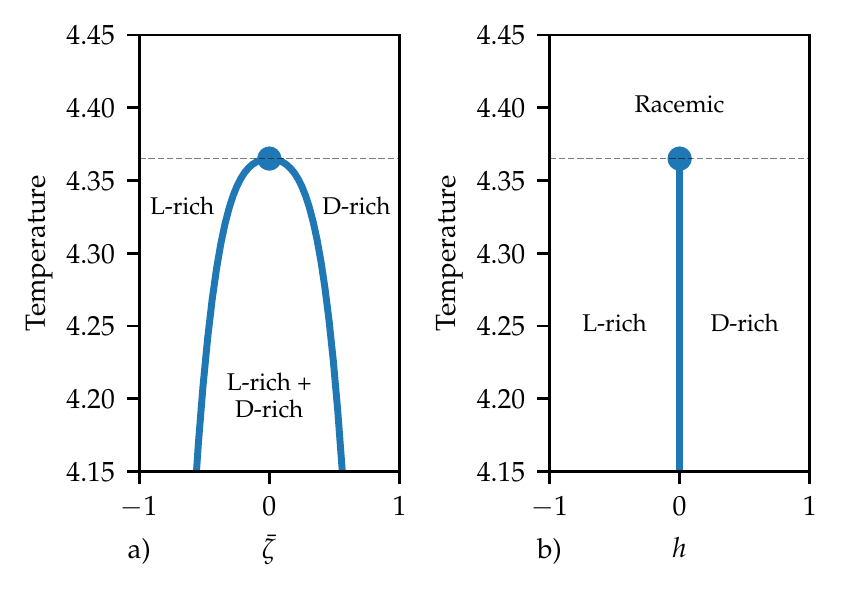}
\caption{\label{fig:FigPhase} Phase diagram of the chiral tetramer model at constant volume and constant number of molecules. a) At constant $\bar\zeta$ and $T$ the phase diagram exhibits a miscibility gap (blue line) between the D-rich and L-rich liquids. b) At constant $h$ and $T$ the phase diagram shows a coexistence line (blue line) between the D-rich and L-rich liquids and ends in a critical point (blue circle).
Note that in our simulations we never observe phase coexistence but, rather, the system spontaneously forms either the L- or D-rich phases, with equal probability.
The critical temperature is also shown with a dashed gray line.
In panel b) the stable phase above the critical point is a racemic mixture.
$\bar\zeta$ is the mean chirality and $h$ is a field conjugate to $\bar\zeta$.}
\end{figure}

The second term in the right-hand side of Eq.~\eqref{eq:pe_field} identifies $N \bar\zeta$ and $h$ as thermodynamic conjugate variables.
The total chirality $N \bar\zeta$ is the extensive variable, in analogy to the volume $V$, the number of particles $N$, or the magnetization $M$, and the field $h$ is the intensive variable, in analogy to the pressure $P$, the chemical potential $\mu$, or an external magnetic field $H$.
We can thus construct phase diagrams for this model while keeping either $N \: \bar\zeta$ or $h$ constant.
The phase diagram in the $\bar\zeta-T$ plane is shown in FIG.~\ref{fig:FigPhase}a and the phase diagram in the $h-T$ plane is shown in FIG.~\ref{fig:FigPhase}b.
In both phase diagrams we consider that the volume and number of molecules is constant.
The $\bar\zeta-T$ phase diagram in FIG.~\ref{fig:FigPhase}a contains a miscibility gap between the D-rich and the L-rich liquids.
It thus strongly resembles the phase diagram of an immiscible mixture (such as oil and water) in the composition-$T$ plane, or the phase diagram of a liquid-gas transition in the $V-T$ plane.
On the other hand, the $h-T$ phase diagram in FIG.~\ref{fig:FigPhase}b exhibits a line of D-rich/L-rich coexistence that ends in a critical point.
The $h-T$ phase diagram is therefore reminiscent of the phase diagram of the liquid-gas transition in the $P-T$ plane, or the phase diagram of the Ising model in the $H-T$ plane.

The simulations reported here correspond to the condition of constant field $h=0$ and $\bar\zeta$ is allowed to change freely ($NVTh$ ensemble).
The phase diagram for this condition is given by FIG.~\ref{fig:FigPhase}b.
We note that we did not simulate the condition of constant $\bar\zeta$ represented in the phase diagram in FIG.~\ref{fig:FigPhase}a.
It would be possible to simulate this system at constant $\bar\zeta$ ($NVT\bar\zeta$ ensemble) by increasing the force constant $k_d$ for the dihedral term in Eq.~\eqref{eq:pe_intra} in order to hinder the interconversion between the L and D enantiomers.
The absence of interconversion (constant $\bar\zeta$) would lead to a phase behavior consistent with FIG.~\ref{fig:FigPhase}a.

We stress that there is a fundamental difference between the fixed-composition binary mixture and the chiral system simulated in this work.
In a fixed-composition binary mixture the system phase-separates and an interface forms between coexisting phases.
In the present chiral system, the composition is not fixed, since each molecule can freely change its chirality.
Thus, below the critical temperature the system avoids the formation of an energetically costly interface ($\lambda > 0$) and stochastically forms either the L-rich or the D-rich mixture, with equal probabilities, and in equilibrium in the thermodynamic limit it does not form coexisting phases.

\begin{figure*}
\includegraphics[width=0.85\textwidth]{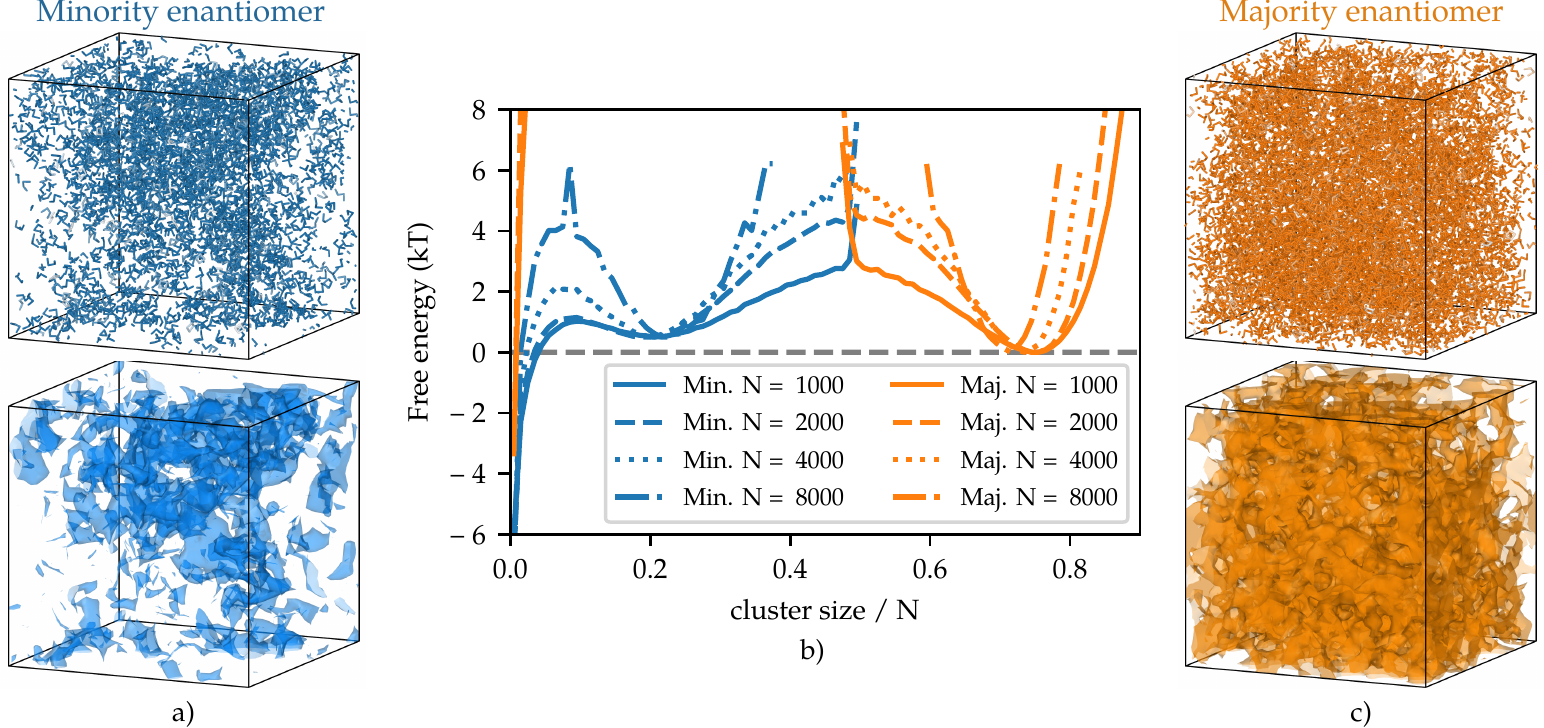}
\caption{\label{fig:structure} Characterization of the structure of the broken-symmetry, subcritical phases at $T=4.3$.  a) Snapshot of the minority enantiomer (D) in the equilibrium L-rich phase for a system of $N=8000$ molecules. In the upper panel molecules are colored according to their chirality (see FIG.~\ref{fig:Fig1} for color scale). In the lower panel a surface mesh is shown in transparent blue to illustrate the cluster distribution. Surface constructed using the alpha-shape method with probe sphere radius of 1.2, as implemented in Ovito\cite{stukowski2009visualization,stukowski2014computational}. b) Free energy vs. cluster size at $T=4.3$ and for different system sizes $N=1000$, 2000, 4000, and 8000. Results are separated into clusters of the majority enantiomer and clusters of the minority enantiomer. c) Snapshot of the majority enantiomer (L) in the equilibrium L-rich phase. In the upper panel molecules are colored according to their chirality (see FIG.~\ref{fig:Fig1} for color scale). In the lower panel a surface mesh is shown in transparent orange to illustrate the cluster distribution.}
\end{figure*}

\subsection{Transformation mechanism and structure of the subcritical phases}

We now turn to characterize the atomistic-level structure of the broken-symmetry, subcritical phases.
Considering that our model favors homochiral interactions, we expect to see clusters of molecules of the same chirality.
For this reason, we calculate the cluster-size distribution $P(n_c)$ of clusters of L-type molecules and of D-type molecules (see section \ref{sec:clustering} for further details).
Due to symmetry considerations, we have combined clustering results of D and L enantiomers.
For visualization purposes we convert the cluster-size distribution $P(n_c)$ to a free energy $F(n_c)=-k_B T \ln P(n_c)$.

The free energy vs.~cluster size $F(n_c)$ is shown in FIG.~\ref{fig:structure}b at $T=4.3$ and for different systems sizes.
The results are separated in clusters of the majority enantiomer (molecules with $\zeta>0$ when $\bar\zeta>0$, or molecules with $\zeta<0$ when $\bar\zeta<0$) and clusters of the minority enantiomer (molecules with $\zeta>0$ when $\bar\zeta<0$, or molecules with $\zeta<0$ when $\bar\zeta>0$).
The free energy curves have three minima that correspond to 1) a cluster of the majority enantiomer of size $n_c/N \approx 0.75$, 2) a cluster of the minority enantiomer of size $n_c/N \approx 0.2$, and 3) small clusters with $n_c/N < 0.05$ mainly from the minority enantiomer.
In order to visualize these three cluster populations, we show in FIG.~\ref{fig:structure}a and c snapshots of an equilibrium configuration for a system of 8000 molecules in the L-rich phase at $T=4.3$.
In FIG.~\ref{fig:structure}a only molecules with $\zeta>0$ are shown, which correspond to clusters of the minority D enantiomer in the L-rich phase.
We observe a cluster of intermediate size, and several smaller clusters, that correspond to the minima at $n_c/N \approx 0.2$ and $n_c/N < 0.05$, respectively.
On the other hand, in FIG.~\ref{fig:structure}c only molecules with $\zeta<0$ are shown, which correspond to clusters of the majority L enantiomer in the L-rich phase.
Here, we observe a large cluster of the majority enantiomer that matches the minimum at $n_c/N \approx 0.75$ in the free energy vs. cluster size curve.
We also note that the cluster-size distribution is system size dependent, at variance with nucleation phenomena wherein the cluster-size distribution is independent of system size\cite{maibaum2008comment,piaggi2016variational}.

\begin{figure*}
\includegraphics[width=0.95\textwidth]{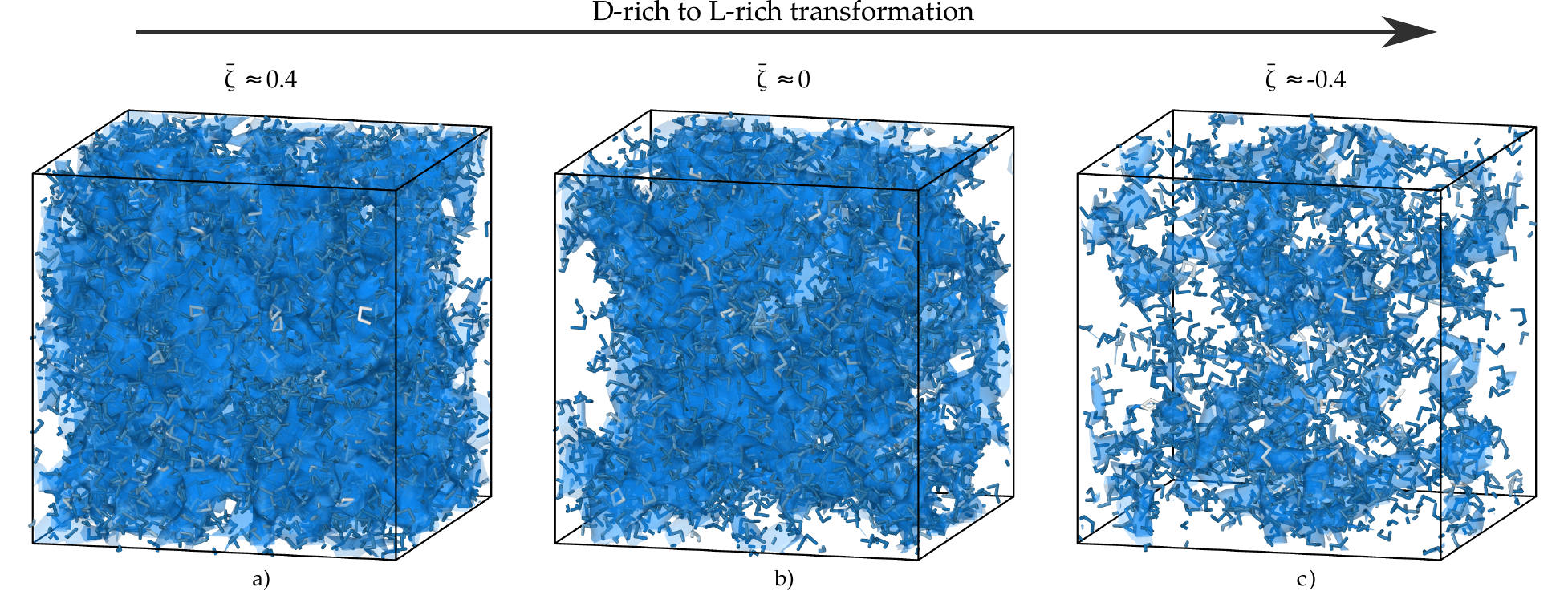}
\caption{\label{fig:transf} Transformation from the D-rich to the L-rich phase. We show snapshots of the configuration of the D enantiomers along the transformation. The snapshots correspond to a) $\bar\zeta \approx 0.4$, b) $\bar\zeta \approx 0$, and c) $\bar\zeta \approx -0.4$. In these snapshots, we display simultaneously the molecules and the surface mesh. Other visualization details are the same as in FIG.~\ref{fig:structure}.}
\end{figure*}

So far, we have focused on the equilibrium configuration of the system.
We now turn to analyze the transition between the D-rich and L-rich phase in a system of $N=4000$ at $T=4.3$.
In FIG.~\ref{fig:transf} we show three snapshots along the spontaneous D-rich to L-rich transformation.
FIG.~\ref{fig:transf}a depicts the majority enantiomer (D) in the D-rich phase with $\bar\zeta \approx 0.4$.
In FIG.~\ref{fig:transf}b we show the D enantiomer at the transition state $\bar\zeta \approx 0$.
The transition state can be associated with the most infrequent cluster $n_c/N \approx 0.5$ in the cluster size distribution in FIG.~\ref{fig:structure}b.
Indeed, the snapshot of the transition state in FIG.~\ref{fig:transf}b shows a cluster of size $n_c/N \approx 0.5$.
Due to symmetry considerations, we expect that the transition state is characterized by a cluster of D with $n_c/N \approx 0.5$ and a cluster of L with $n_c/N \approx 0.5$.
Further analysis also shows that the two clusters with $n_c/N \approx 0.5$ coexist with small clusters of size $n_c/N < 0.01$, probably arising from thermal fluctuations.
The fact that enantiomers form clusters of size $n_c/N \approx 0.5$ at the transition state indicates that the free energy is surface dominated, and lends further credence to our previous finding that the free energy barrier scales like $\Delta F^\ddag \propto N^{2/3}$. 
Lastly, in FIG.~\ref{fig:transf}c we show the minority enantiomer (D) in the L-rich phase, which is the end state of the transformation.

\section{Conclusions}

Our simulations confirm the existence of a second-order symmetry-breaking phase transition in the chiral tetramer model proposed by Latinwo et al.\cite{latinwo2016molecular}~and Petsev et al.\cite{petsev2021effect}
We also show that the critical behavior in this system is compatible with the 3D Ising universality class.
Thus, a similar behavior can be expected in real chiral molecules regardless of the details of the intermolecular interactions.
This symmetry-breaking transition could provide an alternative explanation for the origin of biological homochirality.

Furthermore, our findings indicate that at a slightly subcritical temperature $T\approx0.985T_c$ the free energy barrier $\Delta F^{\ddag}$ for the interconversion between the subcritical D-rich and L-rich liquids increases with system size.
Our data supports a surface-dominated scaling law $\Delta F^{\ddag}\propto N^{2/3}$ where $N$ is the number of molecules.
Also, the increase in free energy barrier with size implies that once a liquid with strong homochiral interactions is sufficiently large, fluctuations between the L-rich and D-rich phases are suppressed.
The system can thus remain in a liquid state highly-enriched in one of the two enantiomers.
Our analysis also shows that in this system there is no driving force for nucleation a result of the equal chemical potentials of the subcritical phases.
We conjecture that nucleation may be triggered through an appropriate chiral external field that creates a chemical potential imbalance.

Finally, we analyze the molecular structure of the subcritical liquids at temperature $T\approx0.985T_c$ using the cluster-size distribution for the majority and minority enantiomers.
We find that the majority enantiomer forms a relatively large cluster of size $\sim 0.75 \: N$.
The minority enantiomer forms a cluster of size $\sim 0.2 \: N$ and smaller clusters of size $< 0.05 \: N$.
In the transformation from the L-rich to the D-rich phase, the transition state is characterized by clusters of size $\sim 0.5 \: N$ for both enantiomers.
The observation of large single clusters for each enantiomer supports the surface-dominated scaling law for $\Delta F^{\ddag}$.

Future work could deal with the calculation of free energy barriers with greater precision in order to provide stronger evidence of the dependence of the free energy barrier $\Delta F^{\ddag}$ with system size.
Enhanced sampling methods, such as umbrella sampling\cite{torrie1977nonphysical}, could be useful to obtain accurate predictions for free energy barriers.
Furthermore, employing more realistic models for the intermolecular interactions could improve our understanding of the relevance of symmetry breaking and liquid-liquid interconversion in real chiral molecules.
Finally, it would be interesting to evaluate critical exponents near the confluence of liquid-liquid and liquid-gas critical points (see ref.~\citenum{wang2022fluid} for a discussion about this phenomenon).


\begin{acknowledgments}
The authors thank Nikolai Petsev and Yiming Wang for assistance in the use of the chiral model implementation and in setting up the simulations.
This work was conducted within the center Chemistry in Solution and at Interfaces funded by the USA Department of Energy under Award DE-SC0019394.
Simulations reported here were performed using the Princeton Research Computing resources at Princeton University which is consortium of groups including the Princeton Institute for Computational Science and Engineering and the Princeton University Office of Information Technology’s
Research Computing department.
\end{acknowledgments}

\section*{Data Availability Statement}

The data that support the findings of this study are openly available in Princeton University DataSpace at \url{https://doi.org/10.34770/aby7-r955}.

\section*{Author Declarations}

\subsection*{Conflict of Interest}

The authors have no conflicts to disclose.

\subsection*{Author Contributions}

\parindent 0em
\textbf{Pablo M. Piaggi:} Investigation (lead); Writing – original draft (lead); Conceptualization (equal); Data curation (lead);  Writing – review \& editing (equal). \textbf{Roberto Car:} Conceptualization (supporting); Supervision (equal); Writing – review \& editing (equal). \textbf{Frank H. Stillinger:} Conceptualization (equal); Supervision (supporting); Writing – review \& editing (equal). \textbf{Pablo G. Debenedetti:} Conceptualization (lead); Supervision (lead); Writing – review \& editing (equal).
\parindent 1em

\section*{References}

%
%

\end{document}